\documentclass[
 reprint,
 amsmath,amssymb,
 aps,
letterpaper,accepted=2025-07-30]{quantumarticle}

\pdfoutput=1

\usepackage{graphicx}
\usepackage{hyperref}
\usepackage[linesnumbered, ruled, vlined, titlenotnumbered, noend]{algorithm2e}
\usepackage{amsmath, amsfonts, amssymb, amsthm, color}
\usepackage{tikz-cd}
\usepackage[capitalise]{cleveref}
\crefname{algocf}{alg.}{algs.}
\Crefname{algocf}{Algorithm}{Algorithms}

\newtheorem{proposition}{Proposition}

\newcommand{\ket}[1]{|#1\rangle}

\newcommand{\plog}{{p_{\log}}}

\DeclareMathOperator{\CliNR}{CliNR}

\newcommand{\Prob}{{\mathbb{P}}}

\newcommand{\Z}{{\mathbb{Z}}}

\newcommand{\Pauli}{{\cal P}}
\DeclareMathOperator{\Gr}{Gr}
\newcommand{\Grassmann}[2]{{\Gr(#1, #2)}}
\DeclareMathOperator{\Gs}{Gs}
\newcommand{\Searchgraph}[2]{{\Gs(#1, #2)}}
\DeclareMathOperator{\opt}{opt}
\DeclareMathOperator{\Proxy}{Proxy}
\newcommand{\proxycost}[1]{\Proxy(#1)}
\DeclareMathOperator{\tabu}{tabu}
\DeclareMathOperator{\cand}{cand}

\newcommand{\registerInput}[0]{R_{\text{in}}}
\newcommand{\registerResource}[1]{R_{\text{res}}^{(#1)}}
\newcommand{\registerVerification}[0]{R_{\text{verif}}}

\begin{document}

\title{
Optimized Clifford Noise Reduction: Theory, Simulations and Experiments
}

\author{Edwin Tham}
\author{Nicolas Delfosse}
\affiliation{
    IonQ Inc.
}

\date{5th August 2025}

\begin{abstract}
We propose several optimizations of the CliNR partial error correction scheme which implements Clifford circuits by consuming a resource state.
Errors are corrected by measuring a sequence of Pauli operators that we refer to as the verification sequence.
We first propose a global optimization algorithm searching for a verification sequence resulting in a low logical error rate using tabu search.
Then, we introduce a proxy for the logical error rate which is easier to evaluate and we design a two-step optimization algorithm.
First, a verification sequence minimizing the proxy is computed, then this sequence is refined by reintroducing the logical error rate.
Finally, we identify a large group of automorphisms of the search space which preserve the proxy and we use this symmetry to reduce the size of the search space.
This results in a $168 \times$ (respectively $20,160 \times$) reduction of the size of the search space for the optimization of verification sequences with three (respectively four) Pauli operators.
Our numerical simulations for 20-qubit Clifford circuits with size $400$ under the ion chain model show that our optimization algorithms improve the performance of CliNR by $25\%$ and that the two-step optimization achieves the same results as the global optimization with $64\%$ fewer evaluations of the logical error rate.
Finally, we perform experiments on a 36-qubit trapped ion quantum computer, without mid-circuit measurements, 
showing that the CZNR variant of CliNR is at breakeven.
\end{abstract}

\maketitle

\section{Introduction}

Large-scale quantum computing applications likely require a fault-tolerant quantum computer in order to correct faults occurring during the computation before they spread to all the qubits~\cite{shor1996fault, preskill2025beyond}. 
However, fault tolerance comes with a massive overhead in term of qubit and gate count.

Partial error correction is likely to play an important role in the near-term as it consumes fewer qubits and fewer gates.
For example, the recent CliNR scheme has a qubit overhead of only about $3\times$ whereas standard quantum error correction codes generally consume tens or hundreds of physical qubits per logical qubit~\cite{delfosse2024low}.
Even though it is not designed to be fault-tolerant, and it does not reduce the logical error rate as much as fault-tolerant quantum error correction, previous simulations show that this scheme could be useful to reduce the logical error rate of quantum circuits in near-term noise regimes.

With CliNR, the Clifford circuit to implement is decomposed into a sequence of $t$ subcircuits and each subcircuit $C$ is implemented in three steps as follows.
\begin{enumerate}
    \item {\bf Resource state preparation.} 
    A resource state $\ket{\psi}$ is prepared on ancilla qubits. This state a stabilizer state~\cite{gottesman1997stabilizer}.
    \item {\bf Resource state verification.} 
    The resource state $\ket{\psi}$ is tested by measuring some of its stabilizers. By definition, these measurements should return a trivial outcome. If not, we know that a fault has occurred, whereupon we reset the qubits and restart the preparation of $\ket{\psi}$.
    \item {\bf Resource state injection.} If all measurement outcomes are trivial, the subcircuit $C$ is applied to an $n$-qubit input state by consuming the resource state $\ket{\psi}$.
\end{enumerate}
This approach to implement quantum gates through resources state has been extensively used before CliNR.
Shor used resource states to perform error correction and logical Toffoli gates in his seminal paper introducing fault-tolerant quantum computation~\cite{shor1996fault}.
Steane and Knill proposed quantum error correction protocols based on different resource states~\cite{steane1997active, knill2005quantum}.
Magic states are resource states that can be consumed to implement non-Clifford gates~\cite{bravyi2005universal}.
Gate teleportation provides a general framework to this approach and it is a core ingredient in CliNR~\cite{gottesman1999quantum}.
Measurement-based quantum computation~\cite{raussendorf2001one, briegel2009measurement} and fusion-based quantum computation~\cite{bartolucci2023fusion} also rely on resource states to perform quantum error correction and fault-tolerant logical operations.
These schemes generally require a long sequence of stabilizer measurements for resource state verification to achieve fault tolerance or to reach the full code distance.
Here, we focus on CliNR and our goal to improve the logical error rate of CliNR implementation of a circuit using a relatively short sequence of stabilizer measurements, which keeps the overhead low.

We refer to the stabilizers selected for the resource state verification as the {\em verification sequence}.
Our main goal is to address the following problem.

\medskip
{\bf The CliNR optimization problem.}
{\em Given a noise model and a Clifford circuit $C$, find a verification sequence that minimizes the logical error rate of the CliNR implementation of $C$.}

\medskip
The main challenge with the CliNR optimization problem is the combinatorial explosion of the number of possible verification sequences.
The number of choices for a $n$-qubit circuit is
\begin{align} \label{eq:count_stabilizer_measurement_sequences}
\prod_{i=0}^{r-1} \left( 2^{2n} - 2^{i} \right)
\end{align}
where $r$ is the verification sequence size.
Therein, we assume that the $r$ stabilizers of the verification sequence are independent.
For $n=10$ and $r=2$ this number is already larger than $10^{12}$ and it reaches more than $10^{36}$ when $n=20$ and $r=3$ which is a relevant regime for CliNR based on previous simulations.

If the circuit is sufficiently small, we can enumerate all possible verification sequences.
Each verification sequence can then be evaluated through Monte-Carlo simulation estimates of the logical error rate of $C$ under the corresponding CliNR implementation.
This provides a solution to the CliNR optimization problem; though, this exhaustive approach is clearly limited to small circuits.

In~\cite{delfosse2024low}, the verification sequence was designed either by selecting stabilizers uniformly at random or by selecting low-weight stabilizers because they can be implemented with fewer gates.
It is natural to expect that a more careful selection of the stabilizers, tailored to the exact noise model and structure of the circuit, would lead to a reduction of the logical error rate achieved with CliNR.

To go beyond these naive optimizations, we introduce a heuristic optimization algorithm (\cref{algorithm:global_optimization}) based on tabu search~\cite{glover1986,glover1989tabu}.
We chose tabu search for its simplicity, but this subroutine can be replaced by other metaheuristic optimization algorithm such as hill climbing, simulated annealing or more involved strategies.
We start from a randomly selected verification sequence.
Then, at each iteration we evaluate a set of neighboring verification sequences by replacing one of the stabilizers within that sequence by a random stabilizer of the resource state.
The neighboring verification sequence that reduces the logical error rate the most is accepted and added to a tabu list -- this prevents the same verification sequence from being revisited over a set number of iterations, and helps the algorithm avoid local minima.

The most resource intensive subroutine in \cref{algorithm:global_optimization} is the cost function evaluation, that is the estimation of the logical error rate of CliNR for a given verification sequence.
To speed up this subroutine, we introduce a proxy cost function, that is an approximation of the logical error rate and we propose a two-step optimization algorithm (\cref{algorithm:two_step_clinr_optimization}).
First, we search for a verification sequence that minimizes our proxy.
Then, in a second step, we refine our verification sequence by introducing the actual cost function.

Our proxy has two features that allow us to speed up the exploration of the search space.
First, it is easier to evaluate than the actual cost function.
Second, it is invariant under the action of a large group of automorphisms of the search space.
This allows us to reduce the size of the search space by identifying points that are equivalent up to the action of this group.
In other words, we reduce the search space by identifying verification sequences which have the same proxy cost.

Interestingly, the resulting reduced search space is a well-studied object in geometry, topology and combinatorics called the {\em Grassmann manifold}, which has applications in machine learning, computer vision or imagine processing~\cite{bendokat2024grassmann}.
In what follows, we use the term {\em Grassmann graph} because we are working with finite fields and then the Grassmann manifold has a graph structure.

We used our two optimization algorithms to improve the CliNR implementation of random Clifford circuits.
We perform numerical simulations for random $n$-qubit Clifford circuits with size $s=n^{1.8}$ and $s=n^{2}$ with $n=20$ under the ion chain model proposed recently as a model for long chains of trapped ions~\cite{ye2025quantum}.
For $s=n^{2}$, the previous CliNR scheme based on a random verification achieves a $37\%$ reduction in logical error rate and the CliNR optimization algorithms proposed in this work achieve an additional $25\%$ reduction.
Moreover, the two-step optimization algorithm (\cref{algorithm:two_step_clinr_optimization}) reaches the same logical error rate as the global optimization algorithm (\cref{algorithm:global_optimization}) using $64\%$ fewer evaluations of the logical error rate.
This reduction in the number of evaluations of the logical error rate is particularly relevant for practical applications where optimizing CliNR may include estimating the logical error rate by executing a program on a real machine.

Finally, we performed experimental demonstrations of the CZNR variant~\cite{delfosse2024low} of CliNR that implements circuits made with CZ gates, showing that the scheme is at breakeven.
These experiments are executed on a IonQ Forte-Enterprise quantum computer with 36 qubits encoded in a long chain of Ytterbium-171 ions.
This machine is not equipped with mid-circuit measurements, therefore we keep the ancilla qubits used for stabilizer measurements idle until the end of the circuit.
We expect that a proper implementation of CZNR with mid-circuit measurements will lead to a further reduction of the logical error rate as it would remove the need to keep qubits idle until the final measurement.

We compare the logical error rate of the direct implementation and the CZNR implementation of 10-qubit circuits with 90 CZ gates and we conclude that our CZNR experiment achieves breakeven performance based on three following observations.
(i) The direct implementation and the CZNR implementation achieve approximately the same logical error rate.
(ii) Increasing the length of the verification sequence keeps the logical error rate constant.
That means that noise introduced by implementing the verification sequence is compensated by the noise it removes.
(iii) Increasing the noise rate in the resource state leads to an increase in the logical error rate.
Observations (i) and (ii) are not enough by themselves to show that the scheme is at breakeven. 
Indeed, one may fulfill these two with a system saturated with noise, {\em i.e.} such the logical error rate remains constant because it has reached a maximum value.
Item (iii) excludes this explanation.
All together, these observations suggest that our CZNR experiment is at breakeven.
A small improvement in gate fidelity may allow us to reach the below-threshold regime where the CZNR implementation has a lower logical error rate than the direct implementation.

The CliNR scheme is related to the coherent parity check (CPC) scheme~\cite{roffe2018protecting, debroy2020extended, van2023single, martiel2025lowoverhead}, which was demonstrated experimentally recently~\cite{van2023single}.
The main advantage of CliNR is that it avoids the exponential blowup in shot count of the CPC scheme.
This is because the Clifford circuit we wish to implement is partitioned into smaller sub-circuits that are implemented through resource states as described above.
Our experiment illustrates this feature with a partition of the circuit into two subcircuits.

In the remainder of the paper, \cref{sec:background} reviews the background and notations and \cref{sec:optimization_problem} provides a detailed description of the CliNR optimization problem.
Our first algorithm that provides an approximate solution of this problem is the global optimization algorithm (\cref{algorithm:global_optimization}) described in \cref{sec:global_clinr_optimization}.
The goal of \cref{sec:two_step_clinr_optimization} is to propose a two-step optimization algorithm based on a proxy for the cost function.
The key theoretical result justifying this algorithm is \cref{prop:search_space_symmetry} which identifies a group of automorphisms of the search space that preserves our proxy.
This proposition also shows that the reduced search space, obtained by identifying the points living in the same orbits under the group action, is the Grassmann graph.
Then, \cref{algorithm:two_step_clinr_optimization} describes the two-step CliNR optimization algorithm which uses the proxy as an intermediate optimization cost function. 
This intermediate step is described in \cref{algorithm:proxy_based_optimization} which is executed in the Grassmann graph instead of the original search space.
\cref{sec:numerics} discusses numerical results comparing our two optimization algorithms. \cref{sec:experiments} presents experimental results on a 36-qubit trapped ion quantum computer showing breakeven performance.

\section{Background}
\label{sec:background}

In what follows, $n \geq 1$ is an integer and $\Pauli_n$ denotes the $n$-qubit {\em Pauli group} quotiented by the phase subgroup $\{\pm I, \pm i I\}$. 
Its elements are called {\em Pauli operators}.
For any vector $u = (u_1, \dots, u_n) \in \Z_2^n$, denote by
$
X^u = X_1^{u_1} \dots X_n^{u_n}
$
the operator of $\Pauli_n$ acting as $X$ on the support of $u$.
The operator $Z^u$ is defined similarly. 
Any operator $P \in \Pauli_n$ is of the form 
$P = X^{u} Z^{v}$ with $u, v \in \Z_2^n$ because its phase can be ignored in the quotiented Pauli group.

Given two Pauli operators $P = X^u Z^v$ and $P' = X^{u'} Z^{v'}$ in $\Pauli_n$, denote 
\begin{align}
    [P, Q] = (u | v') + (u' | v) \pmod 2
\end{align}
where $(x | y) = \sum_{i=1}^n x_i y_i \pmod 2$ is the binary inner product.
The map $[\cdot, \cdot]$ is a symmetric bilinear map, meaning that for all $P, Q$ in $\Pauli_n$, we have $[P, Q] = [Q, P]$, and for all $P, P', Q, Q' \in \Pauli_n$, we have
\begin{align} \label{eq:bilinearity_of_commutation}
[P, QQ'] & = [P, Q] + [P, Q'] \\
[PP', Q] & = [P, Q] + [P', Q] 
\end{align}
where these sums are taken modulo $2$.

We have $[P, Q] = 0$ iff the operators $P$ and $Q$ commute and $[P, Q] = 1$ iff they anti-commute.
Given a subset $S \subseteq \Pauli_n$ of Pauli operators, the set of Pauli operators in $\Pauli_n$ that commute with all the operators of $S$ is denoted by $S^\perp$.
If $S$ generates a rank-$r$ subgroup of $\Pauli_n$,
then $S^\perp$ is a subgroup of $\Pauli_n$ with rank $n-r$.

A {\em Clifford gate} is a unitary gate that maps any Pauli operator onto a Pauli operator by conjugation. Mathematically, a $n$-qubit unitary operation $U$ is a Clifford gate if for all $P \in \Pauli_n$, we have $UPU^{-1} \in \Pauli_n$.
A {\em Clifford circuit} is a sequence of Clifford gates.

A $n$-qubit {\em stabilizer state} is a $n$-qubit state that is fixed by $n$ independent commuting Pauli operators.
Any Pauli operator that fixes a stabilizer state is said to be a {\em stabilizer} of this state.
It is easy to check that Clifford circuits map stabilizer states onto stabilizer states.
As a result, applying a Clifford circuit to a register of qubits initialized in a state $\ket 0$ produces a stabilizer state.

We focus on circuits made with preparation of a qubit in the state $\ket 0$, unitary single-qubit and two-qubit gates and measurements.
For simplicity, we assume that the unitary gates are implemented sequentially.

We consider the standard {\em circuit-level noise} model.
Noise occurring during a preparation, idle step and unitary operation is represented by a depolarizing channel acting on its support, inserted right after the operation. Measurement noise is modeled as a bit-flip of the measurement outcome.
The depolarizing noise rate and the bit-flip rate may depend on the corresponding operation.

Any Pauli error occurring during a $n$-qubit circuit made with qubit preparations, Clifford gates and measurements is equivalent to a Pauli error occurring at the end of the circuit, that we refer to as the {\em output error}. 
The output error can be computed efficiently by conjugating the initial error through the Clifford gates of the circuit~\cite{gottesman1998heisenberg}.
This defines a probability distribution over the set $\Pauli_n$ of all possible output errors, that we call the {\em output noise distribution} of the circuit.

\section{Detailed description of the optimization problem}
\label{sec:optimization_problem}

The CliNR scheme implements a $n$-qubit circuit $C$ using a {\em resource state} $C_R \ket{\Phi_n}$, where
\begin{align}
\ket{\Phi_{n}} = 2^{-n/2} \sum_{x \in \Z_2^n} \ket x \otimes \ket x
\end{align}
is the state of $n$ Bell pairs supported on qubits $i$ and $n+i$ for $i=1,\dots, n$.
The resource state preparation circuit could be $C_R = I \otimes C$ or, alternatively $C_R = C_1^{-1} \otimes C_2$ with $C = C_1 C_2$.

We assume that the stabilizers of the verification are measured sequentially. A stabilizer $P_{i_1} \dots P_{i_w} \in \Pauli_n$ acting on $w$ qubits is measured using a single ancilla qubit prepared in the $\ket +$-state and a sequence of controlled Pauli gates applying $CP_{i_1} \dots CP_{i_w}$ controlled on the ancilla qubit.
The outcome is extracted by measuring the ancilla qubit in the $X$ basis.

\begin{figure*}
    \centering
    \includegraphics[width=.9\linewidth]{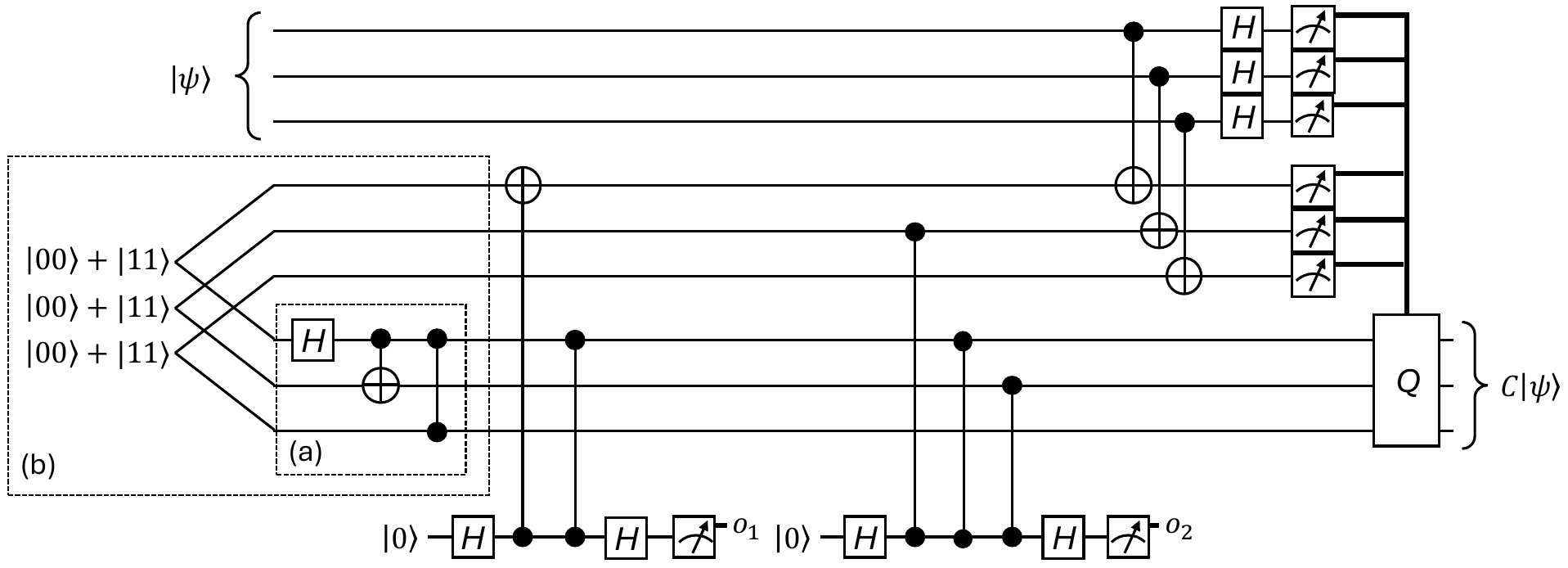}
    \caption{The CliNR implementation of the circuit $C = H_1 CX_{1,2} CZ_{1,3}$.
    The operation $Q$ is a Pauli operator that depends on the measurement outcomes.
    The circuit $C$ appears in (a).
    The resource state is prepared in (b) by applying the circuit $C_R = I \otimes C$ to $n$ Bell states.
    This CliNR implementation uses the verification sequence $V = (XIIZII, IZIZZI)$.
    The measurement outcome $o_1$ is the outcome of $V_1 = XIIZII$ and $o_2$ is the outcome of $V_2 = IZIZZI$.
    If one of these outcomes is non-trivial, the circuit restarts from the beginning of (b).
    }
    \label{fig:clinr_example}
\end{figure*}

The {\em verification sequence} $V$ is defined to be an ordered $r$-tuple of stabilizers of the resource state $C_R \ket{\Phi_n}$.
The CliNR implementation of a circuit $C$ using the verification sequence $V$ is denoted $\CliNR(C_R, V)$.
As an example, \cref{fig:clinr_example} shows the CliNR implementation of the circuit $C = H_1 CX_{1,2} CZ_{1,3}$ with $C_R = I \otimes C$ using the verification sequence 
$V = {XIIZII, IZIZZI}$.

The {\em logical error rate} of the CliNR implementation of a Clifford circuit $C$ is defined to be the probability that the output error is non-trivial, {\em i.e.} $1-\Prob(I)$ where $\Prob$ is the output noise distribution of the CliNR implementation of~$C$. 

The CliNR optimization problem takes as an input a resource state preparation circuit $C_R$ for a Clifford circuit $C$ and the output is a verification sequence $V$ minimizing the logical error rate of $\CliNR(C_R, V)$. 
In what follows, the logical error rate of $\CliNR(C_R, V)$ is denoted $\plog(\CliNR(C_R, V))$.

We optimize the implementation of a single CliNR subcircuit, although CliNR may split the input circuit into subcircuits,
because the verification sequence associated with each subcircuit can be optimized independently.
To further simplify the problem, we can fix the size $r$ of the verification sequence because it is typically small (the proof of Theorem~2 in \cite{delfosse2024low} uses $r = O(\log n)$ for Haar random Clifford circuits acting on $n$ qubits.). 
Therefore, we can run over all $r$ up to a bound $B$, find an optimized sequence for each $r$, and then select the best verification sequence with $r \leq B$.

\section{Global CliNR optimization}
\label{sec:global_clinr_optimization}

This section describes an algorithm for the CliNR optimization problem based on tabu search~\cite{glover1986,glover1989tabu}.
We use tabu search with a random candidate list strategy.
The pseudocode is provided in \cref{algorithm:global_optimization}.

The verification sequence is initialized with $r$ stabilizers of the resource state selected uniformly at random.
Then, we proceed over $i_{\max}$ iterations.
At each iteration, a set of $m$ candidate verification sequences is selected in the neighborhood of the current best verification sequence $V_{\opt}$.
These candidates are obtained by replacing a stabilizer of $V_{\opt}$ by a random stabilizer of the resource state.
After rejecting any sequence that exists in a tabu list, the CliNR logical error rate of the candidate verification sequences is estimated using Monte-Carlo simulations.
The verification sequence $V_{\opt}$ is updated if a lower logical error rate is reached by a candidate verification sequences and this verification sequence is added to the tabu list.
The tabu list is kept to a fixed length $\ell$ by removing entries in first-in-first-out (FIFO) order.

\begin{algorithm}[h!]
\DontPrintSemicolon
\SetAlgoLined
\KwIn{A circuit-level noise model. 
A resource state preparation circuit $C_R$. 
The stabilizer group $S$ of the resource state $C_R \ket{\Phi_n}$.
A positive integer $r$. 
Optimization parameters $\ell$, $m$, and $i_{\max}$.
}
\KwOut{A verification sequence.}
Initialize a verification sequence $V_{\opt}$ with $r$ stabilizers selected uniformly at random in $S$.\;
Initialize an empty tabu list $L_{\tabu}$.\;
\For{$i \gets 0$ \KwTo $i_{\max}-1$}
{
    Initialize an empty candidate list $L_{\cand}$.\;
    Select $j \in [1,r]$ uniformly at random.\;
    \For{$k \gets 0$ \KwTo $m-1$}
    {
        Initialize $V = V_{\opt}$.\;
        Replace $V_j$ by a stabilizer selected uniformly at random in $S$. \;
        Add $V$ to $L_{\cand}$.\;
    }
    \For{$V$ in $L_{\cand}$}
    {
        
        \If{$V$ is not in $L_{\tabu}$}
        {
            Estimate the logical error rate $p_{V} = \plog(\CliNR(C_R, V))$.\;
            \If{$p_{V} < p_{V_{\opt}}$}{
                $V_{\opt} \leftarrow V$.\;
            }
        } 
    }
    \If{$V_{\opt}$ is not in $L_{\tabu}$}
    {
        Append $V_{\opt}$ to $L_{\tabu}$.\;
    }
    \If{$|L_{\tabu}|>\ell$}
    {
        Remove the first element of $L_{\tabu}$.\;
    }
}
\Return{$V_{\opt}$.}\;
\caption{Global CliNR optimization}
\label{algorithm:global_optimization}
\end{algorithm}

Other initializations of this starting verification sequence and different choices for the sequence update are possible. One could for instance favor low-weight operators.
The parameters $\ell$, $m$ and $i_{\max}$ can be chosen through grid search, or any other hyperparameter optimization procedure.

\section{CliNR optimization through a proxy cost function}
\label{sec:two_step_clinr_optimization}

This section describes a two-step optimization algorithm based on a proxy cost function for the CliNR optimization problem.
The proxy cost function is defined in \cref{subsec:proxy_cost_function}.
In \cref{subsec:symmetry_of_search_space}, we identify a group of symmetries of the search space that keep the proxy cost function invariant and we construct the quotient of the search space by this group of symmetry.
\cref{subsec:proxy_cost_optimization} describes an algorithm to optimize the proxy cost function in the quotiented search space.
We refer to this algorithm as the proxy cost optimization.
Our two-step CliNR optimization algorithm, described in \cref{subsec:two_step_clinr_optimization}, first executes the proxy cost optimization and then refine the choice of the verification sequence based on the actual cost function.
The main advantage of this approach is that the proxy cost function can be evaluated without a Monte Carlo simulation and the quotiented search space is significantly smaller than the original search space.
Our simulations presented in \cref{sec:numerics} shows that the two-step CliNR optimization produces better results than the global CliNR optimization.

\subsection{Proxy cost function}
\label{subsec:proxy_cost_function}

The {\em cost function} in the CliNR optimization problem is the logical error rate $\plog(\CliNR(C_R, V))$.

To define our proxy cost function, consider the resource state preparation circuit.
It is a sequence of $s$ operations $O_{i}$ with noise rate $p_i$ in the circuit-level noise model introduced in \cref{sec:background}.
By definition of this noise model, each operation $O_i$ is followed by a random Pauli error $E$ with probability $p_i$ and $E$ is selected uniformly at random in the set of non-trivial Pauli operators acting on the support of $O_i$.
Define $\Omega_i$ to be the set of all possible Pauli operators obtained by propagating a Pauli error $E$ that may occur right after $O_i$ to the end of the circuit.
The propagation of an error can be computed efficiently using Gottesman-Knill algorithm~\cite{gottesman1998heisenberg}. Stim is a convenient tool to perform this computation~\cite{gidney2021stim}.
If the operations $O_i$ acts on $w$ qubits, then the set $\Omega_i$ contains $4^w-1$ errors and each of them has probability $\tilde p_i := p_i / (4^w-1)$.

With these notations, define the {\em proxy cost function} as
\begin{align}
    \proxycost{V} 
    = \sum_{i=1}^s \sum_{P \in \Omega_i}  
    \tilde p_i 
    \delta_{\{P \in V^\perp \backslash S^\perp \}}\cdot
\end{align}
Therein, the resource state preparation circuit is fixed and this proxy assigns a cost with each verification sequence $V$.
The condition $P \in V^\perp$ means that $P$ commutes with all the element of the verification sequence and therefore it is not detected by the verification sequence.
The condition $P \notin S^\perp$ means that $P$ has a non-trivial effect on the resource state and therefore induces a logical error.
This proxy is a first order approximation of the logical error rate when no fault occurs during the resource state verification and injection.

Our motivation for introducing this proxy is twofold. 
First, it is easy to evaluate without a Monte-Carlo simulation.
We precompute the sets $\Omega_i$ and we use them to evaluate the proxy cost $\proxycost{V}$ of a verification sequence $V$ in a single pass.
Second, it is invariant under a large group of symmetry which we describe next.
These symmetries allow us to reduce the size of the search space.

\subsection{Symmetry of the search space}
\label{subsec:symmetry_of_search_space}

The search space has a natural graph structure that we formally introduce now. We also review the definition of the Grassmann graph~\cite{bendokat2024grassmann} and we establish a connection between the two spaces. 
Throughout this section, $S$ is the stabilizer group of the resource state and $r$ is the size of the verification sequence.

The {\em search graph}, denoted $\Searchgraph{S}{r}$, is defined to be the graph whose vertices are the verification sets $V$ containing $r$ independent stabilizers of $S$.
Two vertices of $\Searchgraph{S}{r}$ are connected by an edge if they differ by a single stabilizer, {\em i.e.} $V$ and $V'$ are connected if $|\{i \in [1,r] \ | \ V_i \neq V_i' \}| = 1$.
This definition is motivated by the fact that each modification of the verification in \cref{algorithm:global_optimization} corresponds to a move from a vertex of the search graph to one of its neighbors.

The {\em Grassmann graph}, denoted $\Grassmann{S}{r}$, is defined to be the graph whose vertices are the subgroups of $S$ with rank $r$. Two vertices of $\Grassmann{S}{r}$ are connected if the intersection of the corresponding subgroups is a subgroup with rank $r-1$.

To establish a connection between the search graph and the Grassmann graph, we introduce the map
\begin{align}
\alpha: GL_r(\Z_2) \times \Pauli_n^r 
    & \longrightarrow \Pauli_n^r \\
(M, V)
    & \longmapsto V'
\end{align}
where $GL_r(\Z_2)$ is the group of invertible $r \times r$ matrices with binary coefficients.
The image of $(M, V)$ is the tuple $V' = (V_1', \dots, V_r')$ with  
\begin{align}
V_i' = \prod_{j=1}^r V_j^{m_{i,j}}
\end{align}
where $M = (m_{i,j})_{\substack{1 \leq i, j \leq r}}$.

The following proposition shows that the map $\alpha$ induces a group action on the vertex set of the search graph which allows us to quotient this vertex set producing a reduced search space. The proof of this result in included in \cref{appendix:proof_of_symmetry}.

\begin{proposition}
\label{prop:search_space_symmetry}
The map $\alpha$ induces a group action of the group $GL_{r}(\Z_2)$ onto the vertex set of the search graph $\Searchgraph{S}{r}$ which preserves the proxy cost function.
Moreover, the quotient of the vertex set of the search graph by $GL_{r}(\Z_2)$ is the vertex set of the Grassmann graph $\Grassmann{S}{r}$.
\end{proposition}

The proof of this proposition relies on the fact that $\alpha(M, \cdot)$ is a bijection of the vertex set of the search graph.
In general, it is not a graph automorphism because it does not preserve the edges of the search graph and the Grassmann graph is not the quotient of the search graph by $GL_{r}(\Z_2)$.
We only show that there is a bijection between the vertex set the search graph quotiented by $GL_{r}(\Z_2)$ and the vertex set of the Grassmann graph. 
This bijection is sufficient for our purpose of reducing the size of the search space.

\subsection{Proxy cost optimization}
\label{subsec:proxy_cost_optimization}

This section describes the proxy cost optimization which is the core subroutine of the two-step CliNR optimization algorithm. 
The proxy cost optimization exploits the symmetry of the search space demonstrated in \cref{prop:search_space_symmetry}.

\cref{algorithm:proxy_based_optimization} follows closely the steps of \cref{algorithm:global_optimization} with three major differences.
First, the estimation of the logical error rate is replaced by the proxy cost function.
Second, instead of exploring the search graph, \cref{algorithm:proxy_based_optimization} explores the Grassmann graph to find a verification sequence that reaches a low value for the proxy cost function.
Third, it returns a subgroup $\langle V \rangle$ instead of a verification sequence $V$.
We refer to such a subgroup as a verification subgroup.

For our simulations in \cref{sec:numerics}, we added the condition that candidate subgroups $V$ have rank exactly $r$ in \cref{algorithm:proxy_based_optimization} to our implementation of this algorithm.
To keep the pseudo code simple, we omit this in \cref{algorithm:global_optimization} and \cref{algorithm:proxy_based_optimization}.

\begin{algorithm}[h!]
\DontPrintSemicolon
\SetAlgoLined
\KwIn{A circuit-level noise model. 
A resource state preparation circuit $C_R$.
The stabilizer group $S$ of the resource state.
A positive integer $r$. 
Optimization parameters $\ell$, $m$, and $i_{\max}$.
}
\KwOut{A verification subgroup.}
Initialize a verification sequence $V_{\opt}$ with $r$ stabilizers selected uniformly at random in $S$.\;
Initialize an empty tabu list $L_{\tabu}$.\;
\For{$i \gets 1$ \KwTo $i_{\max}$}
{
    Initialize an empty candidate list $L_{\cand}$.\;
    Select $r-1$ operators $V_1, \dots, V_{r-1}$ in $\langle V_{\opt} \rangle$ uniformly at random.\;
    \For{$j \gets 1$ \KwTo $m$}
    {
        Select a stabilizer $V_r$ in $S \backslash \langle V_{\opt} \rangle$ uniformly at random.\;
        Add $V=(V_1,\dots, V_r)$ to $L_{\cand}$.\;
    }
    \For{$V$ in $L_{\cand}$}
    {
        \If{$V$ is not in $L_{\tabu}$}
        {
            Compute the proxy cost function $\proxycost{V}$.\;
            {\bf if} $\proxycost{V} < \proxycost{V_{\opt}}$ {\bf then} $V_{\opt} \leftarrow V$.\;
        }
    }
    \If{$\langle V \rangle$ is not in $L_{\tabu}$}{
    Append $\langle V \rangle$ to $L_{\tabu}$.\;}
    \If{$|L_{\tabu}|>\ell$}
    {
        Remove the first element of $L_{\tabu}$.\;
    }
}
\Return{$\langle V_{\opt} \rangle$.}\;
\caption{Proxy cost optimization}
\label{algorithm:proxy_based_optimization}
\end{algorithm}

\subsection{Two-step CliNR optimization}
\label{subsec:two_step_clinr_optimization}

We now have all the ingredients to describe the two-step CliNR optimization algorithm.

\cref{algorithm:proxy_based_optimization} returns a subgroup $\langle V \rangle$.
Then, we extract a verification sequence by selecting elements from this subgroup.
The resulting two-step optimization of CliNR is described in \cref{algorithm:two_step_clinr_optimization}.

When the verification subgroup is sufficiently small, the second step could be performed by selecting a verification sequence inside the verification subgroup that minimizes the actual cost function by exhaustive search.

If the verification subgroup is too large for exhaustive search, we performe the second step by executing \cref{algorithm:global_optimization}
to search over the verification sequence inside $\langle V \rangle$, that is replacing $S$ by $\langle V \rangle$ in this algorithm.
This results in \cref{algorithm:two_step_clinr_optimization} which is generally faster than executing \cref{algorithm:global_optimization} with $S$ because $r$ is typically small.

For simplicity, in \cref{algorithm:two_step_clinr_optimization}, we use the same values for the optimization parameters used when calling \cref{algorithm:proxy_based_optimization} and \cref{algorithm:global_optimization}.
Relaxing this assumption could help to further improve the performance of this algorithm.

\begin{algorithm}[h!]
\DontPrintSemicolon
\SetAlgoLined
\KwIn{A circuit-level noise model. 
A resource state preparation circuit $C_R$.
The stabilizer group $S$ of the resource state.
A positive integer $r$.
Optimization parameters $\ell$, $m$, and $i_{\max}$.
}
\KwOut{A verification sequence.}
Compute a verification subgroup $\langle V \rangle$ using \cref{algorithm:proxy_based_optimization} with parameters $\ell$, $m$, and $i_{\max}$.\;
Return a verification sequence obtained by running \cref{algorithm:global_optimization} with $S = \langle V \rangle$ and with parameters $\ell$, $m$, and $i_{\max}$.
\caption{Two-step CliNR optimization}
\label{algorithm:two_step_clinr_optimization}
\end{algorithm}

\subsection{Reduction of the search space}
\label{subsec:reduction_of_the_search_space}

\cref{prop:search_space_symmetry} highlights a key advantage our proxy cost function (together with its low-complexity evaluation). Its large group of symmetry allows us to reduce the size of the search space by identifying the points that differ in a symmetry preserving the proxy cost function.
\cref{prop:search_space_symmetry} proves that the search graph can then be replaced by the Grassmann graph whose number of vertices is given by 
\begin{align}
\label{eq:count_stabilizer_subgroups}
\frac{
    \prod_{i=0}^{r-1} \left( 2^{2n} - 2^{i} \right)
}{
    \prod_{i=0}^{r-1} \left( 2^{r} - 2^{i} \right)
} \cdot
\end{align}
This number is significantly smaller than the number of vertices of the search graph computed in \cref{eq:count_stabilizer_measurement_sequences}.
For $r=3$, this provides a $168\times$ reduction of the size search graph and for $r=4$ the Grassmann graph is $20,160\times$ smaller than the search graph.

In practice, this reduction of the search space allows the two-step CliNR optimization algorithm to reach a better logical error rate than the global CliNR optimization algorithm using the same number of iterations as we observe in our numerical simulations.

\section{Numerical Demonstration}
\label{sec:numerics}

In this section, we apply \cref{algorithm:global_optimization} and \cref{algorithm:two_step_clinr_optimization} to optimize the CliNR implementation of random Clifford circuits and we estimate their logical error rate through numerical simulations.
These numerical results show that the CliNR optimization algorithms proposed in this paper achieve a lower logical error rate than the standard version of the CliNR scheme~\cite{delfosse2024low}.

We estimate the logical error rate of random Clifford circuits and their CliNR implementations under the {\em ion chain model}~\cite{ye2025quantum}.
With this model unitary operations are implemented sequentially. 
Preparations, measurements and resets can be performed simultaneously on an arbitrary subset of qubits.
Each operation is followed by depolarizing noise on the support of the operation with rate $p$ for two-qubit gates, $p/10$ for single-qubit operations, and measurement outcomes are flipped with probability $p/10$.
Noise on an idle qubit during an operation is represented by depolarizing noise with rate $p/100$, except if the qubit is idle during a measurement. 
Then, its noise rate is $\tau_m p / 100$ for some constant $\tau_m$, representing the measurement time, which we set to $\tau_m = 30$.
In the simulations presented in this section, we use $p=10^{-4}$.

\begin{figure*} 
    \begin{centering}
    \includegraphics[width=18cm]{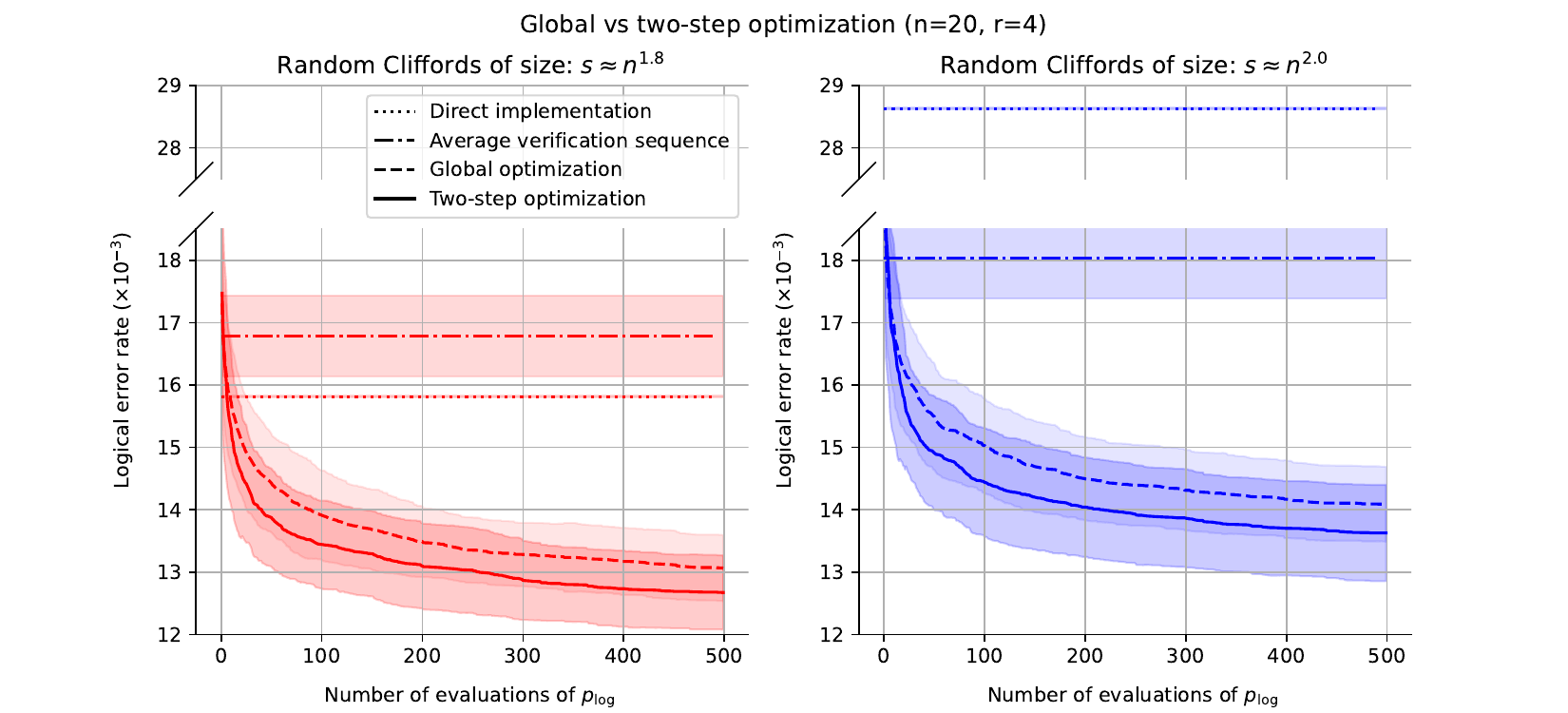}
    \par\end{centering}
    \caption{Comparison of the direct implementation, the CliNR implementation with a uniform random verification sequence proposed in~\cite{delfosse2024low} and the CliNR implementation with verification sequences optimized with global optimization (\cref{algorithm:global_optimization}) and two-step optimization (\cref{algorithm:two_step_clinr_optimization}).
    The standard error in logical error rate is shown in the shaded region around each curve.
    \label{fig:numerical_results}}
\end{figure*}

In \cref{fig:numerical_results}, we show simulation results for two randomly-chosen Clifford circuits $C$; each is chosen by first generating a $n$-qubit Haar-random Clifford circuit with $n$ compiled into the \{$H$, $S$, $S^{\dagger}$, $CX$\} gateset~\cite{Bravyi21-CliffordSynth}, and then truncating it to sizes $s=n^{1.8}$ and $s=n^2$ respectively.
We then compare the logical error rate of $C$ across four different implementations:
(i) the direct implementation of $C$, that is without CliNR (horizontal dotted line), 
(ii) the CliNR implementation of $C$ with a uniform random verification sequence with size $r$ as original proposed in~\cite{delfosse2024low} (horizontal dot-dashed),
(iii) the CliNR implementation of $C$ with a verification sequence with size $r$ optimized with \cref{algorithm:global_optimization} (dashed curve),
(iv) the CliNR implementation of $C$ with a verification sequence with size $r$ optimized with \cref{algorithm:two_step_clinr_optimization} (continuous curve).
These simulations are performed with $n=20$ and $r=4$ and for both (iii) and (iv) we use the values $\ell=10$ and $m=5$ for the parameters of \cref{algorithm:global_optimization,algorithm:two_step_clinr_optimization}and the number of iterations $i_{\max}$ varies.

We plot the logical error rate of the CliNR implementation of $C$ as a function of the number of evaluations of the logical error rate $\plog$, up to a maximum of $500$ evaluations (evaluations of the proxy cost function introduced in \cref{subsec:proxy_cost_function} constitutes a minuscule fraction of optimization, and is not included).
Each curve represents an average over many repetitions of each respective circuit and optimization algorithm, while the shaded area around them indicates one-standard-error boundaries.
For a given circuit $C$, implementation, and optimization procedure, we repeat that procedure identically, 250 times.
Each evaluation of the logical error rate is done over $50,000$ Monte-Carlo shots.
Further, to exclude trivial outcomes, we impose additional constraints not explicit specified in pseudocodes for \cref{algorithm:global_optimization,algorithm:two_step_clinr_optimization}: that the rank of any subgroup $\langle V\rangle$ returned by \cref{algorithm:proxy_based_optimization} have rank $r$, and that trivial entries $I^{\otimes n}$ cannot be admitted into a verification sequence $V$ returned by \cref{algorithm:global_optimization}.

We observe that the optimization algorithms we proposed significantly improve the performance of CliNR. 
In the case of $s=n^{2}$, while random verification sequences yield an average of $\approx 37 \%$ reduction in noise over direct implementation, our global (and two-step) optimization yields a further $\approx 21\%$ ($25 \%$) reduction in logical error rate.
More notably, to achieve the logical error rate reached by global optimization across the full stabilizer group with $500$ evaluations of the cost function, the two-step algorithm required on average only $180$ evaluations.

A similar result is observed in the case of $s=n^{1.8}$.
However, especially noteworthy in that case, is that whereas CliNR with random verification sequences fail to outperform direct implementation, our optimizations yield improvements over the latter of $\approx 17\%$ and $\approx 20\%$ respectively.
This illustrates a case where our optimization algorithms can bring us from above-breakeven to below-breakeven.
We note that $n = 20$ and $r = 4$ for which the search space is already too vast for exhaustive search, with more than $10^{48}$ possible verification sequences.
Indeed, even within an optimization subgroup $\langle V\rangle$ of rank $r=4$, a total of 50,625 possible non-trivial verification sequences is far greater than the $500$ $\plog$ evaluations we used and proves challenging in practice for exhaustive search.

\section{Experimental Demonstration}
\label{sec:experiments}

As a proof-of-concept intended to evaluate suitability of CliNR as a viable partial fault-tolerance noise-reduction technique in the short term, we also proceeded to implement several circuits on IonQ's Forte-Enterprise trapped-ion system.
That system uses hyperfine states of a Ytterbium-171 ion as qubits, in a configuration with up to 36 qubits with all-to-all two-qubit gate connectivity.

We consider the CZNR version~\cite{delfosse2024low} of CliNR which implements a $n$-qubit Clifford circuit $C$, with $n=10$, comprised purely of $CZ$ gates using an resource state of $n$ qubits (which is a graph state in this case) instead of $2n$ for the standard CliNR scheme.

We partition the 36 qubits of the machine into separate functional registers: 
(i) a 10-qubit input register $\registerInput$, onto which $C$ is to be applied; 
(ii) two 10-qubit resource state registers $\registerResource{1}$ and $\registerResource{2}$, where the relevant resource states are prepared; 
(iii) a 6-qubit verification register $\registerVerification$ used to measure the verification sequence.

Instead of an immediate restart upon failure of a verification sequence as in the standard description of CliNR and CZNR, we execute the entire CZNR circuit to completion whereupon measurement of $\registerVerification$ will indicate whether a restart would have been necessary -- the overall CZNR error rate is inferred by discarding such instances.
Such an implementation obviates mid-circuit measurement and hardware reset of qubits.

Define the 10-qubit circuit 
\begin{align}
C = \prod_{1 \leq i < j \leq 10} CZ_{i,j}
\end{align}
which applies a $CZ$ gate to every pair of qubits (45 in total).
The experiment executes the circuit $C^2$ using CZNR with a random product stabilizer state as an input.
It proceeds thus (we direct interested readers to \cref{appendix:ExperimentalDetails} for further details):
\begin{enumerate}
    \item $C_\text{prep}$, a product of $n$ Haar-random single-qubit Clifford gates is chosen. The input state, $C_\text{prep}\ket{0}$ is prepared on $\registerInput$.
    \item The resource state, defined as $C \ket{+}^{\otimes 10}=C H^{\otimes 10}\ket{0}^{\otimes 10}$, is prepared on $\registerResource{1}$ and again on $\registerResource{2}$.
    \item Verification sequences are appended to $\registerResource{1}$ and $\registerResource{2}$, with qubits from $\registerVerification$ being used for readout.
    \item The circuit $C$ is applied twice to $\registerInput$ by consuming the resource states in $\registerResource{1}$ and $\registerResource{2}$. After this, the state $C^2 C_\text{prep}\ket{0}$ resides on $\registerResource{2}$.
    \item $C_\text{prep}^{\dagger}$ is applied to $\registerResource{2}$.
    \item All qubits are measured in the $Z$ basis.\label{step:ExpProcedure}
\end{enumerate}

The fact that a CZ circuit is involutive means that the procedure above performs $C^2 = I$, the identity, on the state $C_\text{prep}\ket{0^n}$.
In an ideal machine, the expected output after Step~\ref{step:ExpProcedure} is the all zero bitstring $\ket{0^n}$.

We ran each circuit in our experiment over 2,048 shots and retained per-shot data for analyses.
Since our experiments do not rely on mid-circuit measurements and feedforward, restarts are accounted for in post-processing (shots with non-trivial stabilizer measurements are discarded, with the discarded fraction being used to infer restart rates).
Similarly, post-teleportation Pauli corrections are realized as a bit-flips in measured bit-strings in post-processing. (See \cref{appendix:ExperimentalDetails} for details).

\begin{figure}[h]
    \begin{centering}
    \includegraphics[width=8cm]{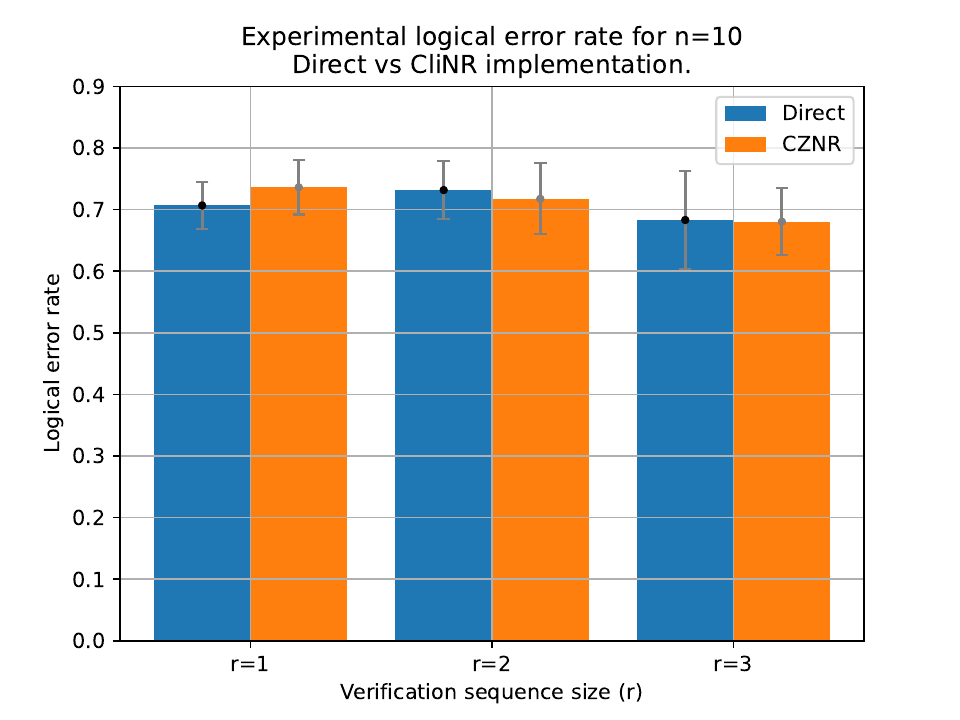}
    \par\end{centering}
    \caption{Experimental logical error rate comparing direct implementation (blue) with CZ noise-reduction or CZNR (orange) circuits implementating the same computation. In the three different experiments, verification sequence size ($r$) is increased, with no significant change in the logical error rate.\label{fig:exp}}
\end{figure}

\begin{figure}
    \begin{centering}
    \includegraphics[width=8cm]{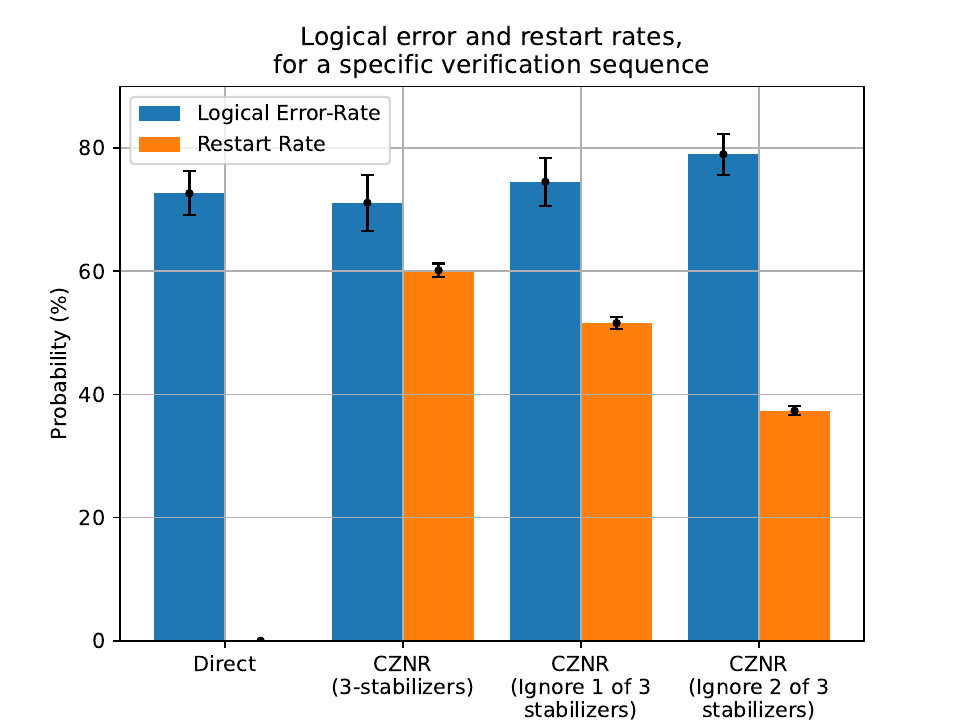}
    \par\end{centering}
    \caption{Comparison of the logical error rate of the CliNR implementations of the same Clifford circuit using a subset of the verification sequence. 
    Restart rates shown here is inferred although the experiment is performed without mid-circuit measurement or restart.
    \label{fig:extrachecks}}
\end{figure}

\cref{fig:exp} shows comparisons of the logical error rates between direct versus CliNR implementations of $C$.
We note that CZNR circuits exhibit comparable logical error rates to circuits under direct implementation, even though the former contains up to $2n(r+1)$ additional two-qubit gates (e.g. 80 additional two-qubit gates, or $\approx 90\%$ more, for $r=3$) than the direct implementation.
We also note that even as $r$, the size of the verification sequence, increases (with a corresponding increase in gate count in the CZNR implementation), the logical error rate remains constant.
A fraction of the noise is detected, manifesting as increased restart rate.

Further, to rule out noise-saturation -- wherein all circuits are maximally noisy -- as a reason for logical error rates remaining flat across implementations with different $r$, we implemented a circuit with $r=3$, and then computed the logical error rate by progressively ignoring some outcomes of the verification sequence.
If the measured stabilizers have any efficacy in detecting faults, then progressively neglecting them should increase logical error rates.
Indeed, in \cref{fig:extrachecks} we observe exactly this -- with the removal of each stabilizer in the verification sequence, the logical error rate worsens while the rejection/restart rate improves.

We \emph{stress} that these results reflect circuit and device performance \emph{\textbf{as is}} with no additional post-processing, filtering, or error-mitigation applied.
Together, these provide evidence that CliNR as implemented on current IonQ's Forte Enterprise is at a breakeven noise threshold.

In these experiments, we note that the optimization techniques for verification sequences we described were not applied.
To make these techniques applicable, one must first refine our cost function to integrate a more precise noise model reflective of hardware, for which previous benchmarking results could provide relevant insights~\cite{chen2024benchmarking}.

\section{Conclusion}

In this work, we designed optimization algorithms to improve the performance of CliNR and its CZNR variant~\cite{delfosse2024low}.
We designed algorithms to explore the search space more efficiently than naive random search and we establish a connection with the Grassmann manifold which allows us to significantly reduce the size of the search space.
We design an optimization algorithm based on tabu search and, through numerical simulations, illustrate the potential gains in CliNR performance for the ion chain model.
Finally, as a proof-of-concept illustrating the potential of CliNR to improve circuit performance in the near term, we also executed CliNR circuits on actual hardware and showed that even on current generation devices, CliNR is already at breakeven performance compared to direct circuit implementation.

Though we applied our ideas here to CliNR specifically, we note that exploiting symmetries of a search space within a stabilizer subgroup can be more widely applicable.
Given that the stabilizer formalism is ubiquitous in quantum information and quantum error correction, we believe that our approach may find other applications in quantum computing. For example, it could be used to optimize stabilizer measurement sequences in quantum error correction as in~\cite{delfosse2020short, tansuwannont2023adaptive}.

\section{Acknowledgments}

The authors thanks John Gamble for his feedback on a preliminary version of this work.

\bibliography{references}

\begin{thebibliography}{25}
\providecommand{\natexlab}[1]{#1}
\providecommand{\url}[1]{\texttt{#1}}
\expandafter\ifx\csname urlstyle\endcsname\relax
  \providecommand{\doi}[1]{doi: #1}\else
  \providecommand{\doi}{doi: \begingroup \urlstyle{rm}\Url}\fi

\bibitem[Bartolucci et~al.(2023)Bartolucci, Birchall, Bombin, Cable, Dawson, Gimeno-Segovia, Johnston, Kieling, Nickerson, Pant, et~al.]{bartolucci2023fusion}
Sara Bartolucci, Patrick Birchall, Hector Bombin, Hugo Cable, Chris Dawson, Mercedes Gimeno-Segovia, Eric Johnston, Konrad Kieling, Naomi Nickerson, Mihir Pant, et~al.
\newblock Fusion-based quantum computation.
\newblock \emph{Nature Communications}, 14\penalty0 (1):\penalty0 912, 2023.
\newblock \doi{https://doi.org/10.1038/s41467-023-36493-1}.

\bibitem[Bendokat et~al.(2024)Bendokat, Zimmermann, and Absil]{bendokat2024grassmann}
Thomas Bendokat, Ralf Zimmermann, and P-A Absil.
\newblock A grassmann manifold handbook: Basic geometry and computational aspects.
\newblock \emph{Advances in Computational Mathematics}, 50\penalty0 (1):\penalty0 6, 2024.
\newblock \doi{https://doi.org/10.1007/s10444-023-10090-8}.

\bibitem[Bravyi and Kitaev(2005)]{bravyi2005universal}
Sergey Bravyi and Alexei Kitaev.
\newblock Universal quantum computation with ideal clifford gates and noisy ancillas.
\newblock \emph{Physical Review A—Atomic, Molecular, and Optical Physics}, 71\penalty0 (2):\penalty0 022316, 2005.
\newblock \doi{http://doi.org/10.1103/PhysRevA.71.022316}.

\bibitem[Bravyi and Maslov(2021)]{Bravyi21-CliffordSynth}
Sergey Bravyi and Dmitri Maslov.
\newblock Hadamard-free circuits expose the structure of the clifford group.
\newblock \emph{IEEE Transactions on Information Theory}, 67\penalty0 (7):\penalty0 4546--4563, 2021.
\newblock \doi{http://doi.org/10.1109/TIT.2021.3081415}.

\bibitem[Briegel et~al.(2009)Briegel, Browne, D{\"u}r, Raussendorf, and Van~den Nest]{briegel2009measurement}
Hans~J Briegel, David~E Browne, Wolfgang D{\"u}r, Robert Raussendorf, and Maarten Van~den Nest.
\newblock Measurement-based quantum computation.
\newblock \emph{Nature Physics}, 5\penalty0 (1):\penalty0 19--26, 2009.
\newblock \doi{https://doi.org/10.1038/nphys1157}.

\bibitem[Chen et~al.(2024)Chen, Nielsen, Ebert, Inlek, Wright, Chaplin, Maksymov, P{\'a}ez, Poudel, Maunz, et~al.]{chen2024benchmarking}
Jwo-Sy Chen, Erik Nielsen, Matthew Ebert, Volkan Inlek, Kenneth Wright, Vandiver Chaplin, Andrii Maksymov, Eduardo P{\'a}ez, Amrit Poudel, Peter Maunz, et~al.
\newblock Benchmarking a trapped-ion quantum computer with 30 qubits.
\newblock \emph{Quantum}, 8:\penalty0 1516, 2024.
\newblock \doi{https://doi.org/10.22331/q-2024-11-07-1516}.

\bibitem[Debroy and Brown(2020)]{debroy2020extended}
Dripto~M Debroy and Kenneth~R Brown.
\newblock Extended flag gadgets for low-overhead circuit verification.
\newblock \emph{Physical Review A}, 102\penalty0 (5):\penalty0 052409, 2020.
\newblock \doi{http://doi.org/10.1103/PhysRevA.102.052409}.

\bibitem[Delfosse and Reichardt(2020)]{delfosse2020short}
Nicolas Delfosse and Ben~W Reichardt.
\newblock Short shor-style syndrome sequences.
\newblock \emph{arXiv preprint arXiv:2008.05051}, 2020.
\newblock \doi{https://doi.org/10.48550/arXiv.2008.05051}.

\bibitem[Delfosse and Tham(2024)]{delfosse2024low}
Nicolas Delfosse and Edwin Tham.
\newblock Low-cost noise reduction for clifford circuits.
\newblock \emph{arXiv preprint arXiv:2407.06583}, 2024.
\newblock \doi{http://doi.org/10.1103/PhysRevLett.134.090603}.

\bibitem[Gidney(2021)]{gidney2021stim}
Craig Gidney.
\newblock Stim: a fast stabilizer circuit simulator.
\newblock \emph{Quantum}, 5:\penalty0 497, 2021.
\newblock \doi{https://doi.org/10.22331/q-2021-07-06-497}.

\bibitem[Glover(1986)]{glover1986}
Fred Glover.
\newblock Future paths for integer programming and links to artificial intelligence.
\newblock \emph{Computers \& Operations Research}, 13\penalty0 (5):\penalty0 533--549, 1986.
\newblock \doi{https://doi.org/10.1016/0305-0548(86)90048-1}.
\newblock Applications of Integer Programming.

\bibitem[Glover(1989)]{glover1989tabu}
Fred Glover.
\newblock Tabu search—part i.
\newblock \emph{ORSA Journal on computing}, 1\penalty0 (3):\penalty0 190--206, 1989.
\newblock \doi{https://doi.org/10.1287/ijoc.1.3.190}.

\bibitem[Gottesman(1997)]{gottesman1997stabilizer}
Daniel Gottesman.
\newblock \emph{Stabilizer codes and quantum error correction}.
\newblock California Institute of Technology, 1997.

\bibitem[Gottesman(1998)]{gottesman1998heisenberg}
Daniel Gottesman.
\newblock The {Heisenberg} representation of quantum computers.
\newblock \emph{arXiv preprint quant-ph/9807006}, 1998.
\newblock \doi{https://doi.org/10.48550/arXiv.quant-ph/9807006}.

\bibitem[Gottesman and Chuang(1999)]{gottesman1999quantum}
Daniel Gottesman and Isaac~L. Chuang.
\newblock Demonstrating the viability of universal quantum computation using teleportation and single-qubit operations.
\newblock \emph{Nature}, 402\penalty0 (6760):\penalty0 390--393, November 1999.
\newblock ISSN 1476-4687.
\newblock \doi{https://doi.org/10.1038/46503}.

\bibitem[Knill(2005)]{knill2005quantum}
Emanuel Knill.
\newblock Quantum computing with realistically noisy devices.
\newblock \emph{Nature}, 434\penalty0 (7029):\penalty0 39--44, 2005.
\newblock \doi{https://doi.org/10.1038/nature03350}.

\bibitem[Martiel and Javadi-Abhari(2025)]{martiel2025lowoverhead}
Simon Martiel and Ali Javadi-Abhari.
\newblock Low-overhead error detection with spacetime codes.
\newblock \emph{arXiv preprint arXiv:2504.15725}, 2025.
\newblock \doi{https://doi.org/10.48550/arXiv.2504.15725}.

\bibitem[Preskill(2025)]{preskill2025beyond}
John Preskill.
\newblock Beyond {NISQ}: {The} {Megaquop} {Machine}.
\newblock \emph{ACM Transactions on Quantum Computing}, 6\penalty0 (3), April 2025.
\newblock \doi{https://doi.org/10.1145/3723153}.

\bibitem[Raussendorf and Briegel(2001)]{raussendorf2001one}
Robert Raussendorf and Hans~J Briegel.
\newblock A one-way quantum computer.
\newblock \emph{Physical review letters}, 86\penalty0 (22):\penalty0 5188, 2001.
\newblock \doi{https://doi.org/10.1103/PhysRevLett.86.5188}.

\bibitem[Roffe et~al.(2018)Roffe, Headley, Chancellor, Horsman, and Kendon]{roffe2018protecting}
Joschka Roffe, David Headley, Nicholas Chancellor, Dominic Horsman, and Viv Kendon.
\newblock Protecting quantum memories using coherent parity check codes.
\newblock \emph{Quantum Science and Technology}, 3\penalty0 (3):\penalty0 035010, 2018.
\newblock \doi{http://doi.org/10.1088/2058-9565/aac64e}.

\bibitem[Shor(1996)]{shor1996fault}
Peter~W Shor.
\newblock Fault-tolerant quantum computation.
\newblock In \emph{Proceedings of 37th conference on foundations of computer science}, pages 56--65. IEEE, 1996.
\newblock \doi{https://doi.org/10.1109/SFCS.1996.548464}.

\bibitem[Steane(1997)]{steane1997active}
Andrew~M Steane.
\newblock Active stabilization, quantum computation, and quantum state synthesis.
\newblock \emph{Physical Review Letters}, 78\penalty0 (11):\penalty0 2252, 1997.
\newblock \doi{https://doi.org/10.1103/PhysRevLett.78.2252}.

\bibitem[Tansuwannont et~al.(2023)Tansuwannont, Pato, and Brown]{tansuwannont2023adaptive}
Theerapat Tansuwannont, Balint Pato, and Kenneth~R Brown.
\newblock Adaptive syndrome measurements for shor-style error correction.
\newblock \emph{Quantum}, 7:\penalty0 1075, 2023.
\newblock \doi{https://doi.org/10.22331/q-2023-08-08-1075}.

\bibitem[van~den Berg et~al.(2023)van~den Berg, Bravyi, Gambetta, Jurcevic, Maslov, and Temme]{van2023single}
Ewout van~den Berg, Sergey Bravyi, Jay~M Gambetta, Petar Jurcevic, Dmitri Maslov, and Kristan Temme.
\newblock Single-shot error mitigation by coherent pauli checks.
\newblock \emph{Physical Review Research}, 5\penalty0 (3):\penalty0 033193, 2023.
\newblock \doi{https://doi.org/10.1103/PhysRevResearch.5.033193}.

\bibitem[Ye and Delfosse(2025)]{ye2025quantum}
Min Ye and Nicolas Delfosse.
\newblock Quantum error correction for long chains of trapped ions.
\newblock \emph{arXiv preprint arXiv:2503.22071}, 2025.
\newblock \doi{https://doi.org/10.48550/arXiv.2503.22071}.

\end{thebibliography}

\appendix

\section{Proof of \cref{prop:search_space_symmetry}}
\label{appendix:proof_of_symmetry}

\begin{proof}
Denote ${\cal V}$ the vertex set of $\Searchgraph{S}{r}$ and consider the restriction of $\alpha$ to the set $GL_r(\Z_2) \times {\cal V}$.
Given that $M$ is invertible, if $V \in {\cal V}$ then $V' = \alpha(M, V)$ is also in ${\cal V}$.
To prove that this maps defined a group action, we need to show that $\alpha(I, V) = V$ and 
$\alpha(M, \alpha(N, V)) = \alpha(MN, V)$.
To prove this result, it is convenient to treat $V$ as a $r \times 2n$ binary matrix that we denote $\bar V$.
Then, the map $\alpha$ can be written as the matrix multiplication $\alpha(M, V) = MV$.
From this description of $\alpha$, the two group properties are immediately clear.

To prove that the proxy is invariant under the group action, it suffices to show that the condition 
$P \in V^\perp \backslash S^\perp$ and $P \in V'^\perp \backslash S^\perp$ are equivalent.
If $P \in V^\perp$, then from \cref{eq:bilinearity_of_commutation} we have 
\begin{align}
[P, V_i'] = \sum_{j=1}^r a_{i,j} [P, V_j] = 0
\end{align}
which proves that $P \in V'^\perp$.
Therefore, we have $V^\perp \subseteq V'^\perp$.
Moreover, we know that $V$ and $V'$ both generate rank-$r$ subgroups, hence $V^\perp$ and $V'\perp$ are subgroups with the same rank $m-r$.
Together with $V^\perp \subseteq V'^\perp$, this proves that $V^\perp = V'^\perp$ and therefore the cost function is invariant under the group action.

Consider the map $\sigma: \Searchgraph{S}{r} \rightarrow \Grassmann{S}{r}$ which maps a vertex $V$ of $\Searchgraph{S}{r}$ onto the vertex $\langle V \rangle$ of $\Grassmann{S}{r}$.
The map $\sigma$ and the quotient map $\pi: \Searchgraph{S}{r} \rightarrow \Searchgraph{S}{r} / GL_r(\Z_2)$ form the following diagram.
\begin{equation}
\begin{tikzcd}
\Searchgraph{S}{r} 
    \arrow[r, "\sigma"] \arrow[d, "\pi"'] 
    & \Grassmann{S}{r} \\
\Searchgraph{S}{r} / GL_r(\Z_2) \arrow[ur, "\bar\sigma"']
\end{tikzcd}
\end{equation}
Therein, we use the notation $\Searchgraph{S}{r}$ and $\Grassmann{S}{r}$ to represent the vertex sets of these two graphs.
The existence of the map $\bar \sigma$ is guaranteed by the fact that $\sigma$ is invariant under the action of the group $GL_r(\Z_2)$.
Moreover, $\sigma$ is surjective as any rank-$r$ group admits admits at least one generating set with $r$ elements.
Therefore, $\bar\sigma$ inherits from the surjectivity of $\sigma$.

To show that $\bar\sigma$ is injective, consider two vertices $V$ and $V'$ of the search graph. If $\bar \sigma(V) = \bar \sigma(V')$ then $V$ and $V'$ generate the same subgroup of $S$.
In particular, each element $V_j'$ of $V'$ can be decomposed into the elements of $V$.
Denote 
\begin{align} \label{eq:proof_injectivity_decomposition_Vjprime}
V_j' = \prod_{k=1}^r V_k^{m_{j,k}}
\end{align}
this decomposition.
Similarly, let 
\begin{align}
\label{eq:proof_injectivity_decomposition_Vi}
V_i = \prod_{j=1}^r V_j'^{n_{i,j}}
\end{align}
be the decomposition of $V_i$ into elements of $V'$.
These decompositions define two $r \times r$ binary matrix $M = (m_{i,j})$ and $N = (n_{i,j})$.
To prove that $V$ and $V'$ are in the same orbit under the action of $GL_r(\Z_2)$, it suffices to show that $M$ is invertible because \cref{eq:proof_injectivity_decomposition_Vjprime} can be written as $V' = \alpha(M, V)$.
Plugging \cref{eq:proof_injectivity_decomposition_Vjprime} inside \cref{eq:proof_injectivity_decomposition_Vi}, we get \begin{align}
\label{eq:proof_injectivity_Vi_equation}
V_i 
= \prod_{j=1}^r \left( \prod_{k=1}^r V_k^{m_{j,k}} \right)^{n_{i,j}} 
= \prod_{k=1}^r V_k^{\sum_{j=1}^r n_{i,j} m_{j,k}}
\end{align}
which holds for all $i=1,\dots,r$.
Therein, we use the fact that the $V_k$ commute which due to the fact that they belong to a stabilizer group $S$.
From \cref{eq:proof_injectivity_Vi_equation}, we obtain
\begin{align}
\sum_{j=1}^r n_{i,j} m_{j,k} = \delta_{i,k}
\end{align}
for all $i, k$, because the operators $V_i$ are independent.
In matrix terms, this reads $MN = I$, proving that $M$ and $N$ belong to $GL_r(\Z_2)$.
Therefore $V$ and $V'$ are equal in the quotient $\Searchgraph{S}{r} / GL_r(\Z_2)$.
This conclude the proof of the injectivity of $\bar \sigma$ and which proves that this map is a bijection between $\Searchgraph{S}{r} / GL_r(\Z_2)$ and $\Grassmann{S}{r}$.
\end{proof}

\section{Experimental Details\label{appendix:ExperimentalDetails}}

\subsection{Experimental Circuit and Verification Sequences}

The Clifford circuit $C$ we chose to implement is defined as:
\[
C=\prod_{i<j}CZ_{i,j}
\]
with $1\leq i,j\leq n$.
When applied to the state $\ket{+^n}$, it prepares the fully connected graph state.

In the absence of verification sequence optimization that is hardware-adapted to our machine's noise, we chose lowest- (highest-) weight stabilizers of $C\ket{+}^{\otimes n}$.
These represent verification sequences that require the least (most) additional two-qubit gates to implement respectively.

Low-weight stabilizers have the form $Y_i Y_j$ with $i\neq j$ and $1\leq i,j\leq n$.
When $r>1$, several such low-weight stabilizers are chosen at random with non-overlapping $i$ and $j$.

By contrast, high-weight stabilizers have the form $X_{i}\prod_{j\neq i}^{n}Z_{j}$, with $1 \leq i,j\leq n$.
When $r>1$, several such high-weight stabilizers are chosen at random with non-overlapping $i$.

For both low- and high-weight stabilizers, we perform the experiment with a verification sequence with size $r=1,2,3$.

\subsection{Compilation and Gate Counts}
All circuits are first compiled into the IonQ native gateset.
That gateset is comprised of $G_\pi$, $G_{\pi/2}$, and $ZZ$ gates, defined in terms of their action on the $Z$ eigenbasis as follows:
\begin{align*}
    G_{\pi}^{\phi} & =\left[\begin{array}{cc}
    0 & e^{-i\phi}\\
    e^{i\phi} & 0
    \end{array}\right]\\
    G_{\pi/2}^{\phi} & =\frac{1}{\sqrt{2}}\left[\begin{array}{cc}
    1 & -ie^{-i\phi}\\
    -ie^{i\phi} & 1
    \end{array}\right]\\
    ZZ_{\phi} & =\left[\begin{array}{cccc}
    e^{-i\phi/2}\\
     & e^{i\phi/2}\\
     &  & e^{i\phi/2}\\
     &  &  & e^{-i\phi/2}
    \end{array}\right]
\end{align*}

In that compilation process \emph{\textbf{no}} optimization passes were applied to circuits in either CZNR or direct implementations; this ensures a fair comparison independent of compilation technique.

Post-compilation, the chosen $C$ (comprised of a block of 45 CZ gates) reduces to 45 $ZZ_{\pi/4}$ gates, along with a layer of ``Virtual-Z'' operators that are not implemented.

The layer of Haar-random Clifford single-qubit gates $C_\text{prep}$ that sets the input state and measurement bases is comprised of exactly one layer of $n$ $G_{\pi/2}$ or $n$ $G_{\pi}$ gates.
It is defined as $C_{\text{prep}}=\prod_{q=1}^{n}\left(G_{\pi}^{(0)}\right)^{c_{q}\cdot a_{q}}\left(G_{\pi/2}^{(a_{q}+b_{q}/2)\pi}\right)^{\neg c_{q}}$.
That is, for each qubit $q$ we choose a set of random bits $a_{q},b_{q},c_{q}\in\{0,1\}$; then, if $a_{q}$ and $c_{q}$ are both `$1$', we apply $G_{\pi}^{(0)}$; otherwise, if $c_{q}=1$, we compute $\phi_{q}=(a_{q}+b_{q}/2)\pi$ and then apply $G_{\pi/2}^{\phi_{q}}$ to that qubit.

The layer of $H^{\otimes n}$ at the start of CliNR stabilizer resource state preparation on $\registerResource{1}$ and $\registerResource{2}$ each adds a layer of $n$ $G_{\pi/2}^{\pi/2}$ gates.

Implementing each stabilizer of weight $w$ in the verification sequence, introduces an additional $w$ $ZZ_{\pi/4}$ gates, along with up to two $2w$ $G_\pi$ and $2w$ $G_{\pi/2}$ gates.

Finally, each CliNR teleportation block introduces an additional $n$ $ZZ_{\pi/4}$ gates, along with $2n$ $G_{\pi/2}$ gates.

\subsection{Post-Processing\label{subsec:PostProc}}
In direct implementation of $C$, the logical error rate as shown in \cref{fig:exp,fig:extrachecks} are computed simply as $p_{\text{log}}=1-\text{Prob}(0^{n})$, the latter being the probability of the bitstring ``000...''.

In CZNR implementations, bits arising from stabilizer measurements are first inspected.
All instances where one or more bits from measuring $\registerVerification$ take value `1' are marked as ``failed'' (suppose there are $N_{\text{fail}}$ of these).
The restart rate is \emph{inferred} as: $p_{\text{restart}}=\frac{N_{\text{fail}}}{N_{\text{shots}}}$, with $N_{\text{shots}}=2048$.

Next, we inspect only instances \textbf{\emph{not}} marked as ``failed''.
For each instance, denote with $\vec{b}_{0}$, $\vec{b}_{1}$, $\vec{b}_{2}$ the bitstring obtained from measuring the registers $\registerInput$, $\registerResource{1}$, and $\registerResource{2}$.
We then compute a correction Pauli $P_{\text{corr}}$:
\begin{align*}
P_{\text{corr}} & =C_{\text{prep}}CX^{\vec{b}_{1}\oplus\vec{b}_{0}}CC_{\text{prep}}^{\dagger}\\
 & =X^{\vec{x}}Z^{\vec{z}}
\end{align*}
Finally, we obtain the corrected output bitstring as: $\vec{b}_{\text{CZNR}}=\vec{b}_{2}\oplus\vec{x}$.
Now suppose there are $N_0$ instances in which $\vec{b}_{\text{CZNR}}=0^n$.
We \emph{infer} the logical error rate as:
\[p_{\text{log}}=\frac{N_{0}}{1-N_{\text{fail}}}\]

\subsection{Restart Rates}

\begin{figure}
    \centering
    \includegraphics[width=.9\linewidth]{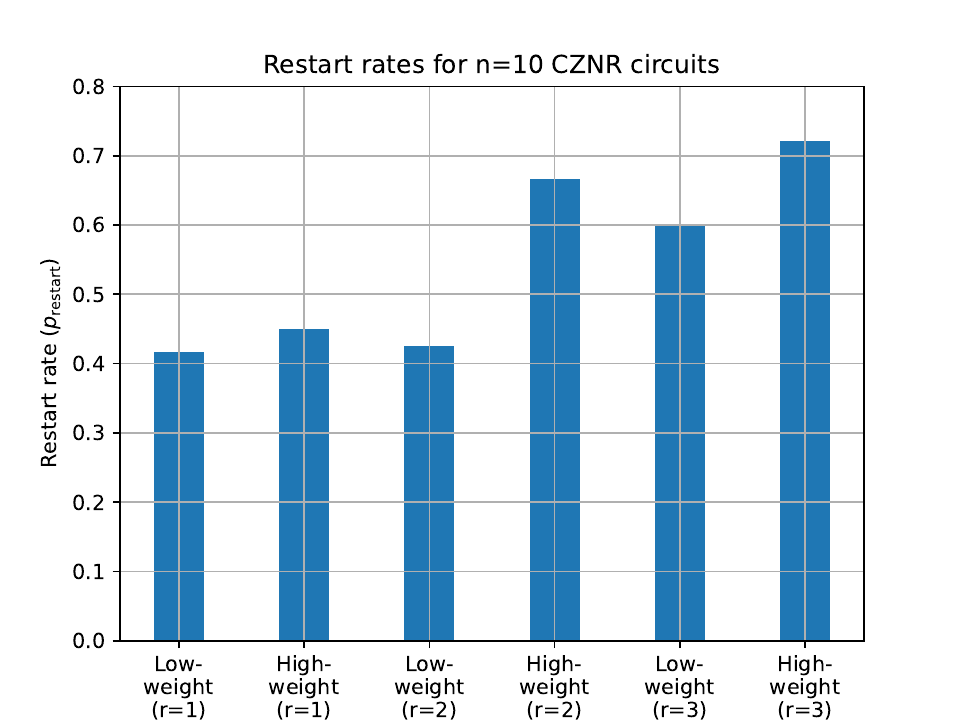}
    \caption{Inferred restart rates (see \cref{subsec:PostProc}) for CZNR circuits.}
    \label{fig:restart}
\end{figure}

In \cref{fig:restart} we show inferred restart rates as computed in \cref{subsec:PostProc}.

When $r>1$, high-weight stabilizers in the verification sequence can lead to higher restart rates than their low-weight counterparts.
Especially, when there are multiple high-weight stabilizer measurements, noise in earlier stabilizer measurements can lead to non-trivial outcomes in later ones, resulting in higher restart rates without corresponding improvement in logical error rates.
We observe this in simulations, and in experimental data.
Here, high-weight stabilizer measurements comprise up to $\approx 27\%$ and $\approx 35\%$ of total two-qubit gates for $r=2$ and $r=3$ respectively).

We also note that, these restart rates reflect the rate at which \emph{any} stabilizer measurement of ancillae in both $\registerResource{1}$ and $\registerResource{2}$ is non-trivial.
A comparable implementation of CliNR in the presence of mid-circuit measurement and restart will yield much higher circuit throughput, with an effective restart rate approximately given by $1-\sqrt{1-p_{\text{restart}}}$ (where $p_{\text{restart}}$ is the restart rate shown here for this experiment).
That distinction lies at the heart of what sets CliNR apart from techniques like the coherent parity check (CPC)~\cite{roffe2018protecting, debroy2020extended, van2023single,martiel2025lowoverhead}.

\end{document}